\documentclass[a4paper]{article}

\usepackage{INTERSPEECH2020}
\usepackage{cite}
\usepackage{subfigure}
\usepackage{graphicx}
\usepackage{url}
\usepackage{multirow}
\usepackage{subfigmat}

\usepackage[linesnumbered,ruled,vlined,commentsnumbered]{algorithm2e}

\SetCommentSty{mycommfont}

\DeclareMathOperator*{\argmax}{arg\,max}
\DeclareMathOperator*{\argmin}{arg\,min}
\newcommand{\pdffigure}[3][width=0.7\linewidth]{
    \begin{figure}[htb]
    \begin{center}
    \IfFileExists{./#2.pdf}{
        \includegraphics[#1]{#2.pdf}
    }{
        \includegraphics[draft]{#2.pdf}
    }
    \end{center}
    \caption{#3}
    \label{fig:#2}
    \end{figure}
}

\title{Neural Speaker Diarization with Speaker-Wise Chain Rule}
\name{
Yusuke Fujita$^1$,
Shinji Watanabe$^2$,
Shota Horiguchi$^1$,
Yawen Xue$^1$,
Jing Shi$^{2,3}$,
Kenji Nagamatsu$^1$
}
\address{
  $^1$Hitachi, Ltd. Research \& Development Group\\
  $^2$Center for Language and Speech Processing, Johns Hopkins University\\
  $^3$Institute of Automation, Chinese Academy of Sciences (CASIA)
  }
\email{yusuke.fujita.su@hitachi.com, shinjiw@ieee.org}

\begin{document}

\maketitle
\begin{abstract}
  Speaker diarization is an essential step for processing multi-speaker audio.
  Although an end-to-end neural diarization (EEND) method achieved state-of-the-art performance, it is limited to a fixed number of speakers.
  In this paper, we solve this fixed number of speaker issue by a novel speaker-wise conditional inference method based on the probabilistic chain rule. In the proposed method, each speaker's speech activity is regarded as a single random variable, and is estimated sequentially conditioned on previously estimated other speakers' speech activities.
  Similar to other sequence-to-sequence models, the proposed method produces a variable number of speakers with a stop sequence condition.
  We evaluated the proposed method on multi-speaker audio recordings of a variable number of speakers.
  Experimental results show that the proposed method can correctly produce diarization results with a variable number of speakers and outperforms the state-of-the-art end-to-end speaker diarization methods in terms of diarization error rate.
\end{abstract}

\section{Introduction}

Speaker diarization is the process of partitioning audio according to the speaker identity, which is an essential step for multi-speaker audio applications such as generating written minutes of meetings \cite{Tranter2006, Anguera2012}.
Related techniques have been evaluated in telephone conversations (CALLHOME \cite{callhome}), meetings (ICSI \cite{Janin03, etin2006OverlapIM}, AMI \cite{Renals2008}), and various {\it hard} scenarios (DIHARD Challenge \cite{Sell2018dihard, Diez2018, Sun2018}). Recent studies on the home-party scenario (CHiME-5 \cite{Barker2018}) reported that
speaker diarization helped improve automatic speech recognition performance \cite{Boeddecker2018, Kanda2018, Kanda2019ICASSP}.

One popular approach to speaker diarization is clustering of frame-level speaker embeddings \cite{Shum2013, Sell2014, Senoussaoui2014, Dimitriadis2017, Romero2017, Maciejewski2018CharacterizingPO, Wang2018LSTM}.
For instance, i-vectors \cite{Dehak2011}, d-vectors \cite{Wan2018}, and x-vectors \cite{Snyder2018} are common speaker embeddings in speaker diarization tasks.
Segment-level speaker embeddings, which are learned jointly with a region proposal network, were also studied \cite{Huang2020}.
Clustering methods commonly used for speaker diarization are agglomerative hierarchical clustering (AHC) \cite{Sell2014, Romero2017, Maciejewski2018CharacterizingPO}, k-means clustering \cite{Dimitriadis2017, Wang2018LSTM}, and spectral clustering \cite{Wang2018LSTM}.
Recently, neural-network-based clustering has been explored \cite{Li2019}.
Although clustering-based methods performed well, it is not optimized to directly minimize diarization errors because clustering is an unsupervised process.
To directly minimize diarization errors in a supervised manner, clustering-free methods have been studied \cite{Zhang2018, Fujita2019E2EDiarization, Fujita2019ASRU}.
%UIS-RNN \cite{Zhang2018} is such an approach that does not cluster speaker embeddings, and that is optimized with a diarization error minimization objective.

End-to-end neural diarization (EEND) \cite{Fujita2019E2EDiarization, Fujita2019ASRU} is one of such clustering-free methods. EEND uses a single neural network that accepts multi-speaker audio and outputs the joint speech activity of multiple speakers. In contrast to the aforementioned methods except for \cite{Huang2020}, EEND handles overlapping speech without using any external module.
The permutation-free training scheme \cite{Hershey2016, Yu2017} and self-attention based network \cite{Vaswani2017} play critical roles in achieving state-of-the-art performance on two-speaker telephone conversation datasets.

However, EEND is limited to a fixed number of speakers because output nodes of the neural network are comprised of multiple speakers' speech activities.
Although one can consider an application to a variable number of speakers by building a neural network that covers a sufficiently large number of speakers, an increasing number of output nodes would require impractical computational resources. 
%However, EEND has only been evaluated on two-speaker audio. In EEND, the number of speakers should be fixed during training and inference. Although one can assume the fixed number to be the maximum number of speakers and can train the model with variable numbers of speakers, it has trouble when estimating the joint speech activity of a large number of speakers at the same time. Due to the speaker label permutation problem becoming more serious, it was hard to apply EEND to the audio of a variable or a large number of speakers.

%To produce a variable and a large number of speakers, in this paper, we propose a speaker-wise conditional decoding method, which is an extension to EEND by placing a speaker-wise conditional decoding module on top of the EEND encoder. In the proposed method, a neural network produces speech activities of one speaker conditioned by the speech activities of previous speakers. According to given multi-speaker audio, the model iteratively produces a variable number of speakers.

In this paper, we solve the fixed number of speaker issue by a novel speaker-wise conditional inference method.
This proposed method regards each speaker's speech activity as a single random variable, and formulates speaker diarization as an estimation of the joint distribution of multiple speech activity random variables.
With the probabilistic chain rule, we can decode a speaker-wise speech activity sequentially conditioned on previously estimated speech activities like the other chain rule-based conditional inference methods, including neural language models \cite{Mikolov2010} and sequence-to-sequence models \cite{Sutskever2014}.
Similar to these conditional inference methods, our method produces a variable number of speakers with an appropriate stop sequence condition.

For training efficiency, we investigate teacher-forcing \cite{Williams1989} like the other sequence-to-sequence models.
The difference between general sequence modeling and our problem is that the order of the speakers is not uniquely determined in advance. Therefore, our approach also searches an appropriate speaker permutation during training to provide the unique speaker order used in teacher-forcing.

The experimental results on CALLHOME and
simulated mixture datasets reveal that our proposed method achieves significant improvement over a conventional EEND method.
Even in a fixed two-speaker configuration, the speaker-wise conditional inference method outperformed a conventional EEND method.
In a variable speaker configuration, the ratio of improvement was more significant in the larger number of speakers. Our source code will be available online at \url{https://github.com/hitachi-speech/EEND}.

\section{Related work}
Our study is inspired by a similar speaker-wise decoding model proposed for speech separation task \cite{Kinoshita2018}, and it was applied in speaker diarization as a down-stream task of speech separation \cite{Neumann2019}. With these methods, the model outputs a ``residual speaker mask'' for the next input. However, the model assumes that the speaker masks are additive and sum to one for each time-frequency bin. This assumption is not directly applicable to our diarization task without using speech separation. 
In this paper, to remove this assumption, we formulate our speaker-wise decoding as a conditional inference with the probabilistic chain rule.
%Bypassing speech separation, our model does not require the clean speech of single speakers.

For a variable number of speakers, AHC \cite{Sell2014} and UIS-RNN \cite{Zhang2018} have been studied.
AHC can generate a variable number of speakers by stopping the cluster merging operation according to a score threshold.
UIS-RNN can detect a new speaker in an online manner by using a Bayesian nonparametric model.
However, these methods fail in processing overlapping speech.
In contrast, our method can handle overlapping speech of a variable number of speakers.

Encoder-decoder based attractor \cite{Horiguchi2020} is another recent approach to the variable number of speakers based on deep attractor networks \cite{Luo2018}.
While this method produces clusters in the embedding space, the proposed method directly produces speech activities using a similar encoder-decoder architecture.

\section{Method}

\begin{figure}[tb]
  \begin{center}
  \subfigure[Conventional EEND method]{
    \includegraphics[width=0.95\linewidth]{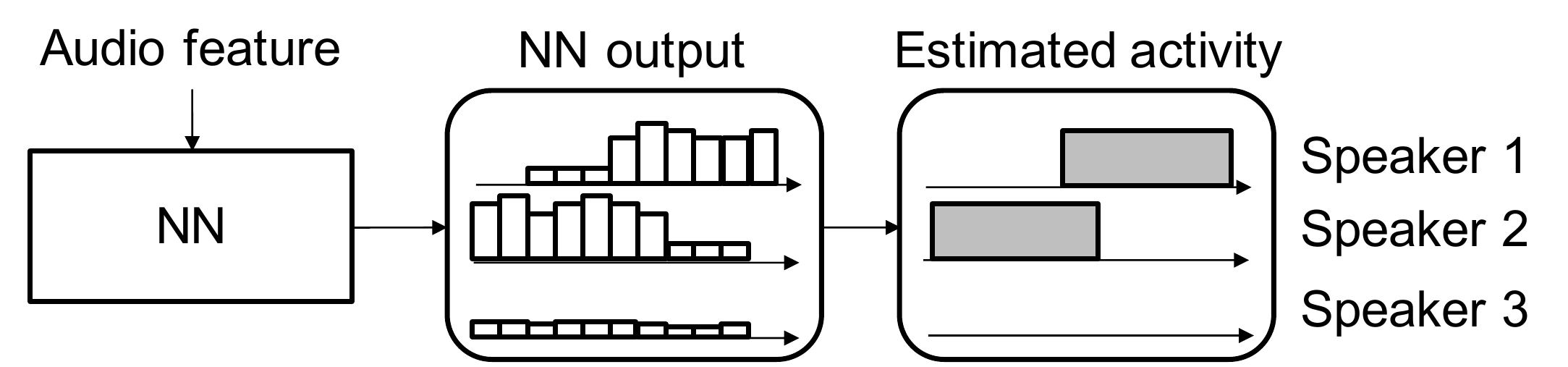}
    \label{fig:eend}
  }
  %\hfill
  \subfigure[Proposed SC-EEND method]{
    \includegraphics[width=0.95\linewidth]{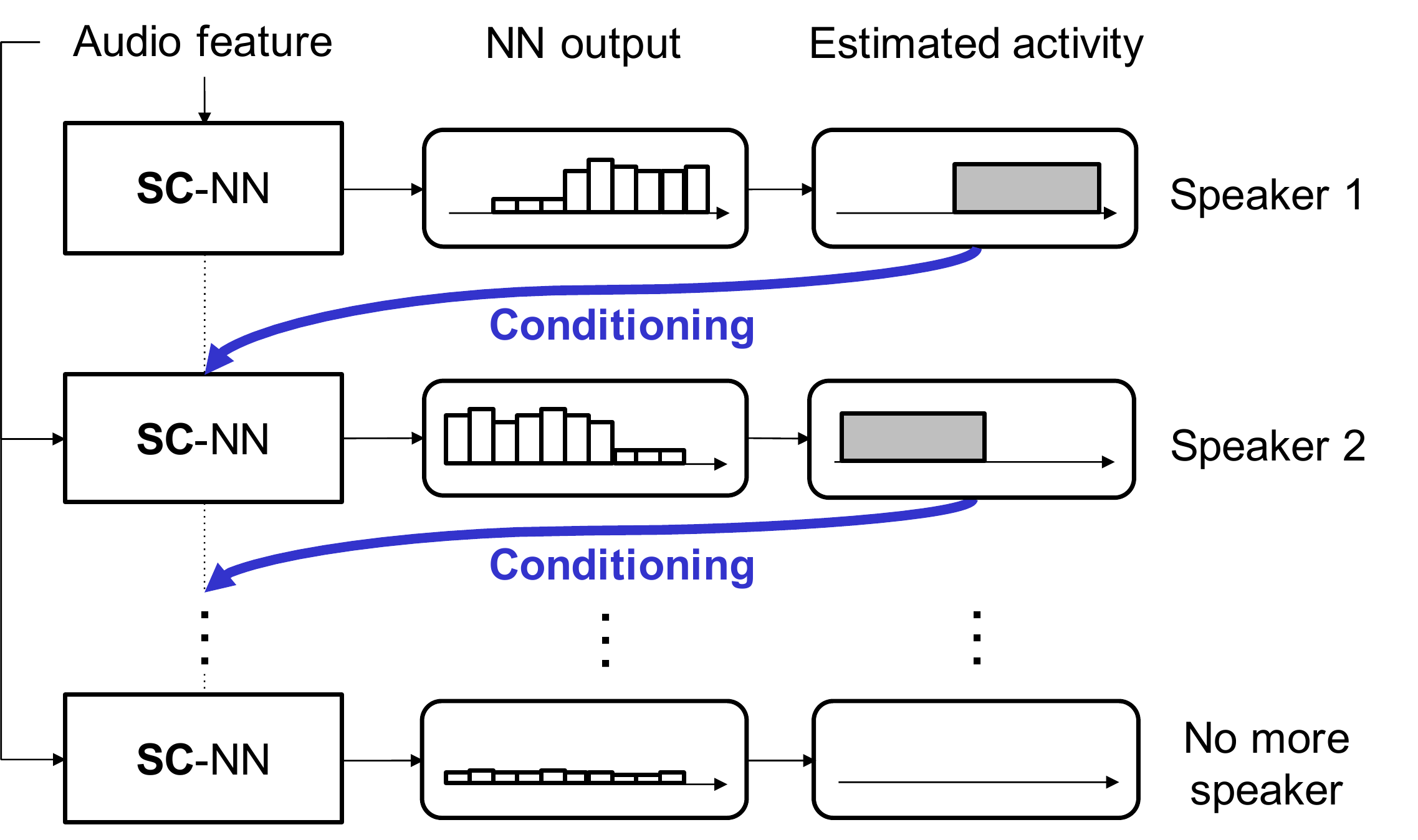}
    \label{fig:csd}
  }
  \end{center}
  \caption{System diagrams of the conventional EEND method and the proposed SC-EEND method.}
\label{fig:systems}
\end{figure}

In this section, we describe speaker-wise conditional EEND (SC-EEND) as an extension of EEND. In the conventional EEND method, the number of speakers should be fixed, as shown in Fig. \ref{fig:eend}.
Instead, the proposed method can produce a variable number of speakers.
As shown in Fig. \ref{fig:csd}, the speaker-wise conditional neural network (SC-NN) produces a speech activity of one speaker conditioned by speech activities of previously estimated speakers. According to given multi-speaker audio, the model can produce a variable number of speakers iteratively.

\subsection{Neural probabilistic model of speaker diarization}
\label{sec:eend}
Given a $T$-length time sequence of $F$-dimensional audio features as a matrix $\mathbf{X} \in \mathbb{R}^{F \times T}$,
speaker diarization estimates a set of speech activities $\mathcal{Y} = \left\{ \mathbf{y}_s \mid s \in \{1,\dots,S \} \right\}$, where $\mathbf{y}_s = \left[y_{s,t} \in \{0,1\} \mid t=1,\dots,T\right]$ is a vector representing a time sequence of speech activity for speaker index $s$, and $S$ is the number of speakers.

In the proposed method, we formulate speaker diarization using a probabilistic model. Our method regards each speaker's speech activity $\mathbf{y}_s$ as a single random variable, and models joint distribution of multiple speech activity random variables for estimating the most probable speech activity $\mathbf{\hat{y}}_s$, as follows:
\begin{equation}
    \mathbf{\hat{y}}_1,\dots,\mathbf{\hat{y}}_S = \argmax_{\mathbf{y}_1,\dots,\mathbf{y}_S} P(\mathbf{y}_1,\dots,\mathbf{y}_S| \mathbf{X}). 
    \label{eq:joint}
\end{equation}

In the EEND method \cite{Fujita2019E2EDiarization}, the joint distribution of multiple speech activity is factorized into speech activity of each speaker using the following conditional independence assumption:
\begin{align}
    P(\mathbf{y}_1,\dots,\mathbf{y}_S| \mathbf{X}) \approx \prod_{s=1}^S P(\mathbf{y}_s| \mathbf{X}).
    \label{eq:independent}
\end{align}
The EEND method models the distribution $P(\mathbf{y}_s| \mathbf{X})$ using a neural network function $\text{NN}()$, which maps an input $\mathbf{X}$ into an output matrix $\mathbf{Z} \in (0,1)^{S \times T}$, as follows:
\begin{equation}
    \mathbf{Z} = \mathrm{NN}(\mathbf{X}). \label{eq:nn}
\end{equation}
$z_{s,t} \in (0,1)$, an element of $\mathbf{Z}$, is interpreted as a posterior of speech activity of speaker $s$ at time index $t$.
Then, $z_{s,t}$ is converted into a binary estimate $\tilde{y}_{s,t} \in \{0,1\}$ using a certain threshold.
Here, the order of speakers (i.e., speaker index $s$) is determined during training.
The training loss for the neural network output $\mathbf{Z}$ is computed as follows,
\begin{align}
    \phi^* &= \argmin_{\phi \in \text{perm}(S)} \sum_{s=1}^S \sum_{t=1}^T \text{BCE}(z_{s,t}, y_{\phi_s,t}) \label{eq:order} \\
    L_\text{PIT} &= \sum_{s=1}^S \sum_{t=1}^T \text{BCE}(z_{s,t}, y_{\phi^*_s,t})
    \label{eq:loss}
\end{align}
where $\text{BCE}(\dot)$ is a binary cross-entropy function between a neural network output and a label, $\text{perm}(S)$ is a set of all possible permutations of a sequence $(1,...,S)$.
The optimal order of speakers is determined as the sequence $\phi^*$.
We refer to the loss function as permutation-invariant training (PIT) loss.

\subsection{Speaker-wise chain rule}

Instead of using the conditional independence assumption in the EEND method, we use a fully-conditional model.
With the probabilistic chain rule, the joint probability in Eq. \ref{eq:joint} is converted into conditional probabilities without using any approximation unlike Eq. \ref{eq:independent}:
\begin{align}
    P(\mathbf{y}_1,\dots,\mathbf{y}_S| \mathbf{X}) = \prod_{s=1}^S P(\mathbf{y}_s|\mathbf{y}_1,\dots,\mathbf{y}_{s-1}, \mathbf{X})
\end{align}
With this model, each speaker's speech activity is sequentially decoded using previously estimated speech activities as conditions.
This model is similar to other conditional inference models.

%With this model, we can decode a speaker-wise speech activity sequentially conditioned on previously estimated speech activities like the other chain rule-based conditional inference methods, including neural language models and sequence-to-sequence models.
%Similar to these conditional inference models, our model produces a variable number of speakers with a stop sequence condition.

%\subsection{Speaker-wise Conditional Neural Network}
In the proposed method, a neural network outputs a vector $\mathbf{z}_s = \left[z_{s,t} \mid t=1,\dots,T\right]$ of speaker index $s$,
\begin{align}
    \mathbf{z}_s &= \text{SCNN}(\mathbf{X}, \mathbf{\tilde{y}}_{s-1}), \label{eq:scnn}
\end{align}
where $\text{SCNN}()$ is a speaker-wise conditional neural network accepts an input $\mathbf{X}$ with a speech activity vector $\mathbf{\tilde{y}}_{s-1} = \left[\tilde{y}_{s-1,t} \mid t=1,\dots,T\right]$ of previous speaker index $s-1$.

To generate a variable number of speakers, Eq. \ref{eq:scnn} is iteratively applied to the next speaker until {\it no speech activity} (i.e. $\mathbf{\tilde{y}}_s$ equals to the all-zero vector) is found.

\subsection{Encoder-Decoder architecture}

Since the proposed speaker-wise conditional neural network generates the output for a variable number of times, the encoder-decoder type of the neural network is a suitable choice.

For the encoder part, similar to EEND \cite{Fujita2019ASRU}, we use the Transformer Encoder \cite{Vaswani2017} as follows:
\begin{align}
    \mathbf{E}_0 &= \text{Linear}^{(F \mapsto D)}(\mathbf{X}) \in \mathbb{R}^{D \times T}, \\
    \mathbf{E}_p &= \text{Encoder}(\mathbf{E}_{p-1}) \in \mathbb{R}^{D \times T} & (1 \le p \le P),
\end{align}
where, $\text{Linear}^{(F \mapsto D)}()$ is a linear projection that maps $F$-dimensional vector to $D$-dimensional vector for each column of the input matrix.
$\text{Encoder}()$ is the Transformer Encoder block that contains a multi-head self-attention layer, a position-wise feed-forward layer, and residual connections. By stacking the encoder $P$ times, $\mathbf{E}_P \in \mathbb{R}^{D \times T}$ is an output of the encoder part.

For the decoder part, the neural network output $\mathbf{z}_s$ for $s$-th iteration is computed as follows:
\begin{align}
    \mathbf{E}'_{s} &= \mathrm{HStack}(\mathbf{E}_P, \mathrm{Linear}^{(1 \mapsto D)}(\mathbf{\tilde{y}}_{s-1})) \in \mathbb{R}^{2D \times T}, \\
    \mathbf{H}_s &= \mathrm{LSTM}^{(2D \mapsto D)}(\mathbf{E}'_{s}, \mathbf{H}_{s-1}) \in \mathbb{R}^{D \times T}, \\
    \mathbf{z}_s &= \sigma(\mathrm{Linear}^{(D \mapsto 1)}(\mathbf{H}_s)) \in (0,1)^{1\times T},
\label{eq:scnn-arch}
\end{align}
where $\mathrm{HStack}()$ concatenates two matrices along with the first axis, $\mathrm{LSTM}^{(2D \mapsto D)}()$ is a uni-directional LSTM that maps $2D$-dimensional vector to $D$-dimensional vector while keeping $D$-dimensional memory cell for each column of the input matrix.
Finally, a linear projection with a sigmoid activation $\sigma()$ produces a $T$-dimensional vector as a neural network output.

\subsection{Teacher-forcing during training}

In Eq. \ref{eq:scnn}, the neural network accepts a speech activity vector of the previous speaker index that is estimated at the previous decoder iteration.
However, the estimation error at the previous iteration hurts the performance at the next iteration.
To reduce the error, we use the teacher-forcing \cite{Williams1989} technique, which boosts the performance by exploiting ground-truth labels.
During training, Eq. \ref{eq:scnn} is replaced with as follows:
\begin{align}
    \mathbf{z}_s^\text{(TF)} &= \text{SCNN}(\mathbf{X}, \mathbf{y}_{s-1}^\text{(TF)}), \label{eq:scnn-tf}
\end{align}
Here, $\mathbf{y}_{s-1}^\text{(TF)}$ is a ground-truth speech activity of speaker index $s-1$.
However, a problem arises with training loss computation in Eq. \ref{eq:order}.
As described in Sec. \ref{sec:eend}, the order of speakers is determined during training. One cannot determine a speaker index $s-1$ before computing the PIT loss, which requires estimates of all speakers.
To alleviate this problem, we examine two kinds of loss computation strategies, as follows.

\subsubsection{Speaker-wise greedy loss}

For each decoding iteration, the optimal speaker index is selected by minimizing binary cross-entropy loss among a set of speaker indices, and the activity of the selected speaker is fed into the next decoding iteration.

\subsubsection{Two-stage permutation-invariant training loss}

Two-stage permutation-invariant training (PIT) loss is computed as Algorithm \ref{alg:twostagepitloss}.
At the first stage, the neural network outputs are computed without teacher-forcing (Eq. \ref{eq:scnn}).
Next, the optimal order of speakers is determined using Eq. \ref{eq:order}.
Then, at the second stage, neural network outputs are computed with teacher-forcing by using the optimal order of speakers. The final loss is computed between the second stage outputs and the ordered labels computed at the first stage.
Note that the computation time of the two-stage process is reasonable since the backward computation is required only in the second stage.
%\footnote{It's worth noting that recomputing PIT loss at the second-stage, i.e. $L' = \min_{\phi \in \text{perm}(S)} \sum_{s=1}^S \sum_{t=1}^T \text{BCE}(z_{s,t}^\text{(TF)}, y_{\phi_s,t})$ as a final loss, didn't work at all, due to the inconsistent orders between teacher and optimal labels.}.

\begin{algorithm}[t]
    \SetAlgoLined
    \DontPrintSemicolon
    \caption{Two-stage PIT loss}
    \label{alg:twostagepitloss}
    \SetAlgoVlined
    \SetKwInOut{Input}{Input}
    \SetKw{In}{in}
    \Input{
        {{$\mathbf{X}$, $\mathcal{Y}$} \tcp*{Audio features and a set of speech activities}}\\
        {{$S_\text{max}$} \tcp*{maximum num. of speakers}}
    } 
    \SetKwInOut{Output}{Output}
    \Output{$L_\text{PIT2}$}
    \BlankLine
    $\mathbf{\tilde{y}}_0 = \mathbf{0}$ \tcp*{Condition for the first iteration}
    \For{$s=1$ \KwTo $S_\text{max}$}{
        $\mathbf{z}_s = \text{SCNN}(\mathbf{X}, \mathbf{\tilde{y}}_{s-1})$, \tcp*{Eq. \ref{eq:scnn}}
        $\mathbf{\tilde{y}}_{s} = [\mathbf{1}(z_{s,t} > 0.5) \mid t = 1,\dots, T]$ \tcp*{Threshold}
    }
    $\phi^* = \argmin_{\phi \in \text{perm}(S)} \sum_{s=1}^S \sum_{t=1}^T \text{BCE}(z_{s,t}, y_{\phi_s,t})$     \tcp*{Optimal order of speakers in terms of PIT loss (Eq. \ref{eq:order})}
    $\mathbf{y}_{0}^\text{(TF)} = \mathbf{0}$ \tcp*{Condition for the first iteration}
    \For{$s=1$ \KwTo $S_\text{max}$}{
        $\mathbf{z}_s^\text{(TF)} = \text{SCNN}(\mathbf{X}, \mathbf{y}_{s-1}^\text{(TF)})$ \tcp*{Eq. \ref{eq:scnn-tf}}
        $\mathbf{y}_{s}^\text{(TF)} = [y_{\phi^*_{s},t} \mid t=1,\dots,T]$  \tcp*{The next condition}
    }
    \tcp{Loss with the optimal order $\phi^*$}
    $L_\text{PIT2} = \sum_{s=1}^S \sum_{t=1}^T \text{BCE}(z^\text{(TF)}_{s,t}, y_{\phi^*_s,t})$ \\
    %\tcp{No speech activity for speaker indices greater than $S$}
    $L_\text{PIT2} \mathrel{+}= \sum_{s=S+1}^{S_\text{max}} \sum_{t=1}^T \text{BCE}(z^\text{(TF)}_{s,t}, 0)$ \tcp*{No speech}
\end{algorithm}

\section{Experimental setup}

\subsection{Data}
\label{sec:data}

We prepared simulated training/test sets for both two-speaker and variable-speaker audio mixtures. We also prepared real adaptation/test sets from CALLHOME \cite{callhome}. % and DIHARD2 \cite{Ryant2019} datasets.
The statistics of the datasets are listed in Table \ref{tab:set}.
For the simulated dataset with a variable number of speakers (Simulated-vspk), the overlap ratio is controlled to be similar among the differing number of speakers. The simulation method is the same as \cite{Fujita2019ASRU}. For the CALLHOME-2spk and CALLHOME-vspk sets, we used the identical set of the Kaldi CALLHOME diarization v2 recipe \cite{Povey_ASRU2011}\footnote{\url{https://github.com/kaldi-asr/kaldi/tree/master/egs/callhome_diarization}}, thereby enabling a fair comparison with the x-vector clustering-based method.
%For DIHARD2, we used the same configuration as Track2 of the second DIHARD challenge \cite{Ryant2019}.
\begin{table}[tb]
\caption{Statistics of training/adaptation/test sets.}
\label{tab:set}
\centering
\begin{tabular}{lrrrr} \hline
 & Num. & Num. & Avg. & Overlap \\
 & spk & rec & dur & ratio \\ \hline
\multicolumn{3}{l}{\bf Training sets} & & \\
\: Simulated-2spk & 2 & 100,000 & 87.6 & 34.4 \\
\: Simulated-vspk & 1-4 & 100,000 & 128.1 & 30.0 \\
\hline
\multicolumn{3}{l}{\bf Adaptation sets} & & \\
\:CALLHOME-2spk & 2 & 155 & 74.0 & 14.0 \\
\:CALLHOME-vspk & 2-7 & 249 & 125.8 & 17.0 \\
%\:DIHARD2-dev& 1-10 & 192 & 443.6 & 9.7 \\
\hline
\multicolumn{3}{l}{\bf Test sets} & & \\
\:Simulated-vspk & 1-4 & 2,500 & 128.1 & 30.0 \\
\:CALLHOME-2spk & 2 & 148 & 72.1 & 13.0 \\
\:CALLHOME-vspk & 2-6 & 250 & 123.2 & 16.7 \\ 
%\:DIHARD2-eval & 1-9 & 194 & 414.4 & 8.9 \\
\hline
\end{tabular}

\end{table}

\subsection{Model configuration}
\subsubsection{x-vector clustering-based (x-vector+AHC) model}
We compared the proposed method with a conventional clustering-based system \cite{Sell2018dihard}, which were created using the Kaldi CALLHOME diarization v2 recipe.
The recipe uses AHC with the probabilistic linear discriminant analysis (PLDA) scoring scheme.
The number of clusters was fixed to be two for the two-speaker experiments, while it was estimated using a PLDA score for the variable-speaker experiments.
% Though the original recipe used oracle speech/non-speech marks, we used the speech activity detection model in Kaldi \footnote{The SAD model: \url{http://kaldi-asr.org/models/m4}}.

\subsubsection{EEND and SC-EEND models}
\label{sec:eendsetup}
We built self-attention-based EEND models and the proposed SC-EEND models, mostly based on the configuration described in \cite{Fujita2019ASRU}. The configurations have small differences between the two-speaker and variable-speaker experiments, as follows.

For the two-speaker experiments, we used four encoder blocks with 256 attention units containing four heads.
For the variable-speaker experiments, we used four encoder blocks with 384 attention units containing six heads. We used a subsampling ratio of 20 for variable-speaker experiments, which is twice larger than that of two-speaker experiments (10).
Note that conventional EEND does not handle a variable number of speakers. We trained a fixed four-speaker model with zero-padded labels for three or fewer speakers in the training data.

%For SC-EEND models, the hyper-parameters of their encoder are the same as the EEND models described above. %For the decoder part, we used the same dimension $D$ with the number of attention units in the encoder part.

\section{Results}

We evaluated the models with the diarization error rate (DER). On the DER computation, overlapping speech and non-speech regions are also evaluated in the experiments. We used a collar tolerance of 250 ms at the start and end of each segment.%except for DIHARD2 evaluations with no collar tolerance.

\subsection{Experiments on fixed two-speaker models}

DERs on the two-speaker CALLHOME are shown in Table \ref{tab:twospk}.
The proposed SC-EEND without teacher forcing (TF) was slightly worse than conventional EEND. With teacher-forcing, DER was significantly reduced and outperformed the conventional EEND method. For the loss computation strategy, two-stage PIT loss (PIT+TF) was slightly better than speaker-wise greedy loss (Greedy+TF).

\begin{table}[tb]
    \caption{DERs on two-speaker CALLHOME.}
    \centering
    %\begin{tabular}{cc|cc} \hline
    %Model & Training & no-adapt & adapted \\ \hline
    %x-vector+AHC & - & N/A & 11.53 \\
    %EEND & PIT & 11.32 & 9.70 \\
    %SC-EEND & PIT & 12.52 & 9.95 \\
    %SC-EEND & Greedy+TF & {\bf 11.09} & 9.01 \\
    %SC-EEND & PIT+TF & 12.46 & {\bf 8.86} \\ \hline 
    %\end{tabular}
    \begin{tabular}{cc|c} \hline
    Model & Training & DER \\ \hline
    x-vector+AHC & - & 11.53 \\
    EEND & PIT  & 9.70 \\
    SC-EEND & PIT & 9.95 \\
    SC-EEND & Greedy+TF  & 9.01 \\
    SC-EEND & PIT+TF & {\bf 8.86} \\ \hline 
    \end{tabular}

    \label{tab:twospk}
\end{table}

\subsection{Experiments on a variable number of speakers}

DERs on the variable-speaker simulated test set are shown in Table \ref{tab:variable}.
For SC-EEND without TF, we observed no significant improvement from the conventional EEND. With TF, again, significant improvement was observed, particularly on a large number of speakers. PIT+TF was significantly better than Greedy+TF.

\begin{table}[tb]
    \caption{DERs on variable-speaker simulated test set.}
    \centering
    \begin{tabular}{cc|cccc} \hline
    & & \multicolumn{4}{c}{Num. of speakers} \\
    Model & Training & 1 & 2 & 3 & 4 \\ \hline
    EEND & PIT & 1.16 & 6.40 & 11.59 & 21.75 \\
    SC-EEND & PIT & 0.96 & 6.32 & 11.75 & 22.52 \\
    SC-EEND & Greedy+TF & 0.85 & 5.25 & 10.56 &  18.28 \\
    SC-EEND & PIT+TF & \bf 0.76 & \bf 4.31 & \bf 8.31 & \bf 12.50 \\ \hline 
    \end{tabular}
    \label{tab:variable}
\end{table}

DERs on the variable-speaker CALLHOME is shown in Table \ref{tab:variablecallhome}.
Even without TF, the SC-EEND outperformed the conventional EEND and x-vector+AHC method. SC-EEND with TF boosted performance significantly. The DER was even better than 17.94\%, which is the DER of x-vector+AHC with the oracle number of speakers.

\begin{table}[tb]
    \caption{DERs on variable-speaker CALLHOME. Note that Greedy+TF adaptation model$\dagger$ was evaluated at $20^{th}$ epoch, because the adaptation was not stable after the epoch.}
    \centering
%    \begin{tabular}{cc|cc} \hline
%    Model & Training & no-adapt & adapted \\ \hline
%    x-vector+AHC & - & N/A & 19.01 \\
%    EEND & PIT & 27.07 & 20.47 \\
%    SC-EEND & PIT & 25.50 & 17.42 \\
%    SC-EEND & Greedy+TF & 23.70 & 18.07$\dagger$ \\
%    SC-EEND & PIT+TF & \bf 23.67 & \bf 15.75 \\ \hline 
%        \end{tabular}
    \begin{tabular}{cc|c} \hline
    Model & Training & DER \\ \hline
    x-vector+AHC & - & 19.01 \\
    EEND & PIT &  20.47 \\
    SC-EEND & PIT &  17.42 \\
    SC-EEND & Greedy+TF & 18.07$\dagger$ \\
    SC-EEND & PIT+TF & \bf 15.75 \\ \hline 
        \end{tabular}
\label{tab:variablecallhome}
\end{table}

\subsection{Analysis on speaker counting}

For the variable-speaker CALLHOME experiment, we analyzed the accuracy of speaker counting.
The results are shown in Table \ref{tab:numberofspeaker}.
The proposed method achieved better speaker counting accuracy than the x-vector+AHC method, while it was still hard to handle more than four speakers.

\begin{table}[htb]
    \caption{Speaker counting results on variable-speaker CALLHOME. SC-EEND models was trained with PIT+TF.}
    \centering
    \begin{subfigmatrix}{2}
    %\resizebox{\linewidth}{!}{
    \setlength{\tabcolsep}{5.1pt}
    \subtable[x-vector+AHC (Acc: 54.6\%)]{
    \begin{tabular}{cc|ccccc} \hline
     & & \multicolumn{5}{c}{Estimated} \\
     & & 2 & 3 & 4 & 5 & 6  \\ \hline
     \multirow{5}{*}{\rotatebox{90}{Reference}} 
     &2 & \bf 84 & 62 & 2 & 0 & 0  \\
     &3 & 18 & \bf 51 & 5 & 0 & 0  \\
     &4 &  2 & 12 & \bf 6 & 0 & 0  \\
     &5 &  0 &  4 & 1 & \bf 0 & 0  \\
     &6 &  0 &  1 & 2 & 0 & \bf 0  \\ \hline 
    \end{tabular}
    }
    \subtable[SC-EEND (Acc: 74.8\%)]{
    \begin{tabular}{c|ccccc} \hline
      & \multicolumn{5}{c}{Estimated} \\
      & 2 & 3 & 4 & 5 & 6  \\ \hline
     %\multirow{5}{*}{Ref} 
     2 & \bf 130 & 17 & 1 & 0 & 0  \\
     3 & 17 & \bf 54 & 3 & 0 & 0 \\
     4 &  4 & 13 & \bf 3 & 0 & 0 \\
     5 &  0 &  3 & 2 & \bf 0 & 0  \\
     6 &  0 &  2 & 1 & 0 & \bf 0 \\ \hline 
    \end{tabular}
    }
    %}
    \end{subfigmatrix}
    \label{tab:numberofspeaker}
\end{table}
%On the variable-number experiment, two-stage PIT loss was significantly better than speaker-wise greedy %loss.
%For emphasizing the difference in detail, learning curves of different models are shown in Fig. %\ref{fig:valid_der}.
%The training curves suggest that Teacher Forcing was affected by the optimal order of speakers. In this regard, PIT loss was more optimal than greedy loss.
%\pdffigure[width=0.9\linewidth]{valid_der}{Learning curves on variable-speaker models. DERs were evaluated on 50-second chunks of Simulated-vspk without a collar tolerance.}

%Finally, the variable-speaker models are tested on DIHARD2 dataset. The results are shown in Table \ref{tab:dihard}. The results were comparable to the state-of-the-art result \cite{Landini2020} and revealed that the proposed method is also applicable to such a hard diarization scenario by employing a neural network adaptation.
%\begin{table}[tb]
%    \caption{DERs on DIHARD2 dataset}
%    \centering
%    \begin{tabular}{c|cc} \hline
%    Model  & DER \\ \hline
%    DIHARD-2 baseline \cite{Ryant2019} & 40.86 \\
%    BUT System \cite{Landini2020, Landini2020ICASSP} & {\bf 27.11} \\ \hline
%    SC-EEND (8kHz) & 34.71 \\
%    SC-EEND (16kHz) & - \\ \hline
%    \end{tabular}
%    \label{tab:dihard}
%\end{table}

\section{Conclusions}

We proposed a speaker-wise conditional inference method as an extension to the end-to-end neural diarization method. Experimental results showed that the proposed method outperformed the conventional EEND method in variable-speaker scenarios. When estimating a larger number of speakers, the proposed method showed its advantage more significantly.
The proposed method achieved better speaker counting accuracy, but it was still hard to handle more than four speakers.
We will explore such hard scenarios, including DIHARD challenges for our future work.
%\newpage
\bibliographystyle{IEEEtran}

\bibliography{mybib}

% \begin{thebibliography}{9}
% \bibitem[1]{Davis80-COP}
%   S.\ B.\ Davis and P.\ Mermelstein,
%   ``Comparison of parametric representation for monosyllabic word recognition in continuously spoken sentences,''
%   \textit{IEEE Transactions on Acoustics, Speech and Signal Processing}, vol.~28, no.~4, pp.~357--366, 1980.
% \bibitem[2]{Rabiner89-ATO}
%   L.\ R.\ Rabiner,
%   ``A tutorial on hidden Markov models and selected applications in speech recognition,''
%   \textit{Proceedings of the IEEE}, vol.~77, no.~2, pp.~257-286, 1989.
% \bibitem[3]{Hastie09-TEO}
%   T.\ Hastie, R.\ Tibshirani, and J.\ Friedman,
%   \textit{The Elements of Statistical Learning -- Data Mining, Inference, and Prediction}.
%   New York: Springer, 2009.
% \bibitem[4]{YourName17-XXX}
%   F.\ Lastname1, F.\ Lastname2, and F.\ Lastname3,
%   ``Title of your INTERSPEECH 2020 publication,''
%   in \textit{Interspeech 2020 -- 20\textsuperscript{th} Annual Conference of the International Speech Communication Association, September 15-19, Graz, Austria, Proceedings, Proceedings}, 2020, pp.~100--104.
% \end{thebibliography}

\end{document}